\documentstyle[aps,prl,multicol,epsf]{revtex}

\def\be{\begin{equation}}
\def\en{\end{equation}}                  
\newcommand{\bi}[1]{\mbox{\boldmath$#1$}}
\newcommand{\av}[1]{\langle{#1}\rangle}
\def\p{\partial}
\def\bea{\begin{eqnarray}}
\def\ena{\end{eqnarray}}
\newcommand{\ppp}[3]{{\bigg(}\frac{\partial {#1}}{\partial {#2}}{\bigg )}_{#3}}

\begin{document}
\draft
\bibliographystyle{prsty}
\title{Dynamic van der Waals Theory 
of two-phase fluids  in heat flow 
}
\author{Akira  Onuki}
\address{Department of Physics, Kyoto University, 
Kyoto 606-8502, Japan}
\date{\today}
\maketitle

\begin{abstract}

We present a dynamic van der Waals theory. 
It is  useful to study phase separation 
 when the temperature varies in space.  
We  show  that if heat flow is applied 
to liquid suspending a gas droplet  at zero gravity,  
a convective flow occurs such that 
the temperature gradient  
within the droplet nearly vanishes. 
As the heat flux is increased, 
the droplet  becomes 
attached  to the heated  wall that is wetted by liquid 
in equilibrium. In one case corresponding to 
partial wetting by gas,  
an apparent contact angle can be defined.  
In the other case with larger heat flux, 
the droplet completely wets   the heated 
wall expelling   liquid.

\end{abstract}

\begin{multicols}{2}

\date{ }
\pagestyle{empty}

In usual theories of 
phase transitions,   the  
fluctuations of the temperature $T$ are assumed to be 
 small and are  neglected. 
However, there can be situations 
in which phase transitions  occur 
in inhomogeneous  $T$.   For example, 
wetting properties near the gas-liquid critical 
point are very sensitive to applied heat flux 
\cite{Beysens,Pomeau} and  boiling processes remain 
largely unexplored \cite{Nikolayev,Physica}. 
To treat such problems we  propose to  start with  
 coarse-grained  entropy 
and energy, from which  the temperature 
is defined as a functional of the order parameter and 
the energy density.

Let us consider one-component fluids, where 
 the order parameter is the number density $n$. 
In the van der Waals theory 
the energy density $e$ 
and the entropy $s$ per particle are given by 
$e= d nk_BT/2- \epsilon v_0n^2$ and 
$s=(d/2) \ln T+ \ln (1/v_0n-1) +{\rm const.}$, 
respectively, in terms of $T$ 
and  $n$ \cite{Onukibook}.  
Here $v_0$ and $\epsilon$ 
represent the molecular  volume  
and the magnitude of the 
attractive  potential, respectively, and 
 $d$ is the space dimensionality. 
To describe situations in the presence of 
interfaces, we generalize this description 
by including  gradient contributions
 to the total entropy and the total internal 
energy as 
\be 
{\cal S}= \int d{\bi r}\bigg [ns-\frac{1}{2}C |\nabla n|^2\bigg ],
\en 
\be 
{\cal E}= \int d{\bi r} \bigg [ e+\frac{1}{2}K |\nabla n|^2 \bigg ]. 
\en 
The gradient terms 
represent  a decrease of the entropy and an increase of the energy 
due to  inhomogeneity of $n$. 
Here  the   internal 
energy density  is written as 
\be 
\hat{e}= e+ K|\nabla n|^2/2, 
\en  
including the gradient part. 
We assume that the coefficients  $C$ and $K$ 
 are  simply  constants  and that 
$s$ in Eq.(1)  depends on 
$n$ and $e$ as  
\be
  s =  k_{\rm B}      \ln [(e/n+ \epsilon v_0n)^{d/2}
  ({1}/{v_0n}-1 )]  +{\rm const.}
\en 
This  relation follows from  the van der Waals theory. 
Our scheme starts with Eqs.(1), (2), and (4), where 
the temerature is  not yet introduced. 
We then define 
 the local temperature $T=T(n,e)$ by  
\be 
\frac{1}{T}= \bigg (\frac{\delta}{\delta e}{\cal S}\bigg )_n
= n\ppp{s}{e}{n}, 
\en 
where $n$ is fixed in the derivatives. 
This definition of $T$ 
 is analogous to that in 
a micro-canonical ensemble.  
We  also define the local  chemical potential 
$\hat \mu$ (per particle) including the gradient 
contributions by 
\be 
\hat{\mu} = -T \bigg ( \frac{\delta {\cal S}}{\delta n}\bigg )_{\hat{e}}= 
\mu - T\nabla \cdot \frac{M}{T}\nabla n ,
\en  
where   
$\hat{e}$ in Eq.(3) is fixed, 
 $\mu= -T(\p (ns)/\p n)_e$, and 
\be 
M=CT+K. 
\en 
Maximization of $\cal S$ under a fixed 
total particle number $\int d{\bi r}n$ 
and a fixed total energy 
$\cal E$ leads to the equilibrium conditions 
$T=$const. and $\hat\mu=$const. 
As first derived by van der 
Waals \cite{Onukibook,vander},
 the equilibrium interface density profile 
$n=n(x)$ is 
determined by $\mu(n,T) -M d^2n/dx^2=\mu_{\rm cx}$ 
with the surface tension $\gamma= M  
\int_{-\infty}^\infty dx  (dn/dx)^2$, 
where  $\mu_{\rm cx}$ is the chemical 
potential on the coexistence curve \cite{surface}.

Next we  set up  the  dynamic equations  \cite{Landau}. Hereafter   gravity 
will be  neglected.  
The mass density $\rho= m n$ 
($m$ being the particle mass) obeys 
$\p \rho/\p t = -\nabla\cdot(\rho{\bi v})$, 
where $\bi v$ is the velocity field. 
The momentum density $ \rho {\bi v}$  obeys   
\be 
\frac{\p}{\p t} \rho{\bi v}= -\nabla\cdot(\rho {\bi v}{\bi v}) 
-\nabla\cdot (\tensor{\Pi}-\tensor{\sigma}). 
\en 
The  $\tensor{\Pi}= 
\{\Pi_{ij}\}$ is  the reversible stress  tensor 
invariant  with respect to time reversal 
${\bi v}\rightarrow -{\bi v}$. The  
 $\tensor{\sigma}= \{\sigma_{ij}\}$ is  the dissipative  stress tensor 
of the form  $\sigma_{ij}= \eta (\nabla_i v_j + \nabla_j v_i) 
+(\zeta-2\eta/d)\delta_{ij}\nabla\cdot{\bi v}$ 
with $\nabla_i= \p/\p x_i$  in 
 terms of   the   viscosities $\eta$ and  $\zeta$.  
The (total) energy density 
$e_{\rm T}=\hat{e}+ \rho {\bi v}^2/2$ including the kinetic part  
 is governed by 
\be 
\frac{\p}{\p t} {e}_{\rm T}=
 -\nabla\cdot[ e_{\rm T}{\bi v} + 
(\tensor{\Pi}-\tensor{\sigma})\cdot{\bi v}]  
+\nabla \cdot (\lambda \nabla T)  ,
\en 
where $\lambda$ is the thermal conductivity. 
We  construct the stress tensor  
$\Pi_{ij}$ such that 
the entropy production rate within  the fluid 
 assumes  the usual   form \cite{Landau},  
\be 
\frac{\p}{\p t} {\cal S}  =
\int d{\bi r}\frac{1}{T} [\nabla\cdot\lambda\nabla T+  
\sum_{ij} \sigma_{ij} \nabla_i v_j]. 
\en 
If there is no heat flow from outside, 
the above quantity becomes non-negative-definite. 
To obtain  Eq.(10) we need  to require the general relation,  
\be 
 \nabla \cdot \bigg (\frac{1}{T}  {\tensor{\Pi}} \bigg ) 
= - n\nabla  \bigg ( \frac{\delta }{\delta n}{\cal S}\bigg )_{\hat e} 
- \hat{e}\nabla \bigg (\frac{\delta }{\delta {\hat e}}{\cal S} \bigg )_{n}.
\en 
Using the van der Waals pressure  
$p=nk_BT/(1-v_0n) -\epsilon v_0n^2$ we then find  
\be 
\Pi_{ij}=  (p+p_1) \delta_{ij} -M\nabla_i n \nabla_j n, 
\en 
where  
$p_1= M (n\nabla^2 n +|\nabla n|^2/2) 
+ Tn \nabla n\cdot\nabla (M/T)$.

As an example,  we  consider 
a  bubble  with radius $R$  in heat flow. 
Thermocapillary bubble migration  has 
long been observed  especially in space 
\cite{Young,space,Beysens1}.  
Theoretically, however, 
the bubble motion has been examined in 
the absence of first-order phase transition 
at the interface such as in the case of air bubbles 
in silicone oil \cite{Young}.  In such cases 
the temerature dependence of the 
surface tension $\gamma$ causes 
a bubble velocity  of order 
$v_{\rm Y}= R
\gamma_1 ({dT}/{dz})_\infty/\eta$  
in a given  temperature 
gradient $(dT/dz)_\infty$ (Marangoni effect),   
where $\gamma_1=-d\gamma/dT>0$ and 
$\eta$ is  the  viscosity in liquid. 
In one-comonent fluids the effect is 
drastically altered due to latent heat 
generation or absorption at the interface \cite{OnukiM}. 
We  point out  that  a convective  velocity   
of order $v_{\rm O} = 
\lambda ({dT}/{dz})_\infty /\rho'T\Delta s$  
inside the bubble is sufficient  to 
carry the applied   heat flux, 
where $\rho'$ is the 
 mass density in gas, $\Delta s$ is the entropy difference 
between gas and liquid per unit mass, and  
 $\lambda$ is the thermal conductivity in liquid.  
Note that the ratio
\be  
M_1=  v_{\rm Y}/v_{\rm O}= 
R\gamma_1\rho'T\Delta s/\eta\lambda
\en  
is a very large number  of order $R/a$ with 
$a$ being a microscopic length. 
With phase change the bubble velocity 
is  of order $v_{\rm O}$ and is very slow \cite{OnukiM}. 
Furthermore,  the temperature 
inside the bubble should be nearly 
homogeneous due to the latent heat convection. 
This is a  natural consequence in one-component fluids without 
contamination \cite{Beysens,OnukiM}, 
because the pressure should become  nearly 
homogeneous outside  the bubble     
and the fluid near the interface  
should be close to the coexistence curve 
in the $p$-$T$ phase diagram.

In our simulation 
we set $K=0$ and 
then  the interface thickness 
is $\ell=(Cv_0/2k_{\rm B})^{1/2}$ far from 
the critical point (since $(\mu-\mu_{\rm cx})/k_{\rm B}T$ 
is of order 1 for $n \sim v_0^{-1}$). 
Space is  measured in units of 
$\ell$ and the system 
length is $L=202\ell$. 
Time is  measured  in units of 
$t_0= m\ell^2/\eta v_0$. 
Here we assume constant viscosities with 
 $\eta=\zeta$. 
The strength of dissipation 
is represented by the normalized 
viscosity $\eta^*= v_0\eta/\ell\sqrt{m\epsilon}$. 
We set $\eta^*=0.1^{1/2}$. 
On all the boundary walls 
we assume \cite{deGennes,Puri}
\be 
n= n_{\rm w}  + \ell_{\rm w} {\bi \nu}\cdot{\nabla n},
\en 
where $\bi \nu$ is the outward normal unit vector 
at the surface. We set $n_{\rm w}=5n_{\rm c}/2$ and 
$\ell_{\rm w}= 2\sqrt{2}\ell$, for which the wall is 
wetted by liquid in equilibrium in our simulation.
We prepared  an  initial   equilibrium state with  
a circular gas bubble 
with radius $50$ placed at the center of the cell. 
The temperature was  
 at $T=0.875T_{\rm c}$, where 
the density is  
$0.37 n_{\rm c}$ in gas and $1.74 n_{\rm c}$ in liquid. 
The bottom  boundary 
($y=0$) was then increased 
by a constant  $\Delta T$, while 
 the top boundary ($y=L$) was 
held at the 
initial temperature  
$T_{\rm t}= 0.875T_{\rm c}$. 
(Here we use "bottom" and "top" though 
no gravity is assumed.) 
The side boundaries ($x=0$ and $L$) 
are   insulating 
(${\bi \nu}\cdot\nabla T=0$).  
Furthermore,  the density-dependence 
of the thermal conductivity is 
crucial away from criticality  \cite{Bird}, so 
we set $\lambda=(v_0\eta/k_{\rm B} m) n$ with the  
thermal diffusivity   $D_T$ being of 
order 1 (in units of $\ell^2/t_0= v_0\eta/m$). 
Then $\lambda$ in liquid is larger than that in gas 
by 5 times.

(i) First, for  very small $\Delta T$, 
 the bubble  position is  
shifted toward the bottom  without  shape changes. 
Fig.1 shows such a steady state  
at $\Delta T= 0.00675T_{\rm c}$ taken at $t=3.4\times 10^3$ 
after application of heat flux. 
Here the migration   is   stopped by   the  
 repulsion from the  walls 
due to the wetting condition in Eq.(14). 
The velocity field $\bi v$ 
is almost parallel to the interface, 
but there is a  small velocity  component 
through the interface. Thus, at the interface,  
 first-order phase transition  
 takes place and latent heat is generated or absorbed. 
As demonstrated in Fig.2, 
  latent heat transport is so  efficient 
 that   the temperature gradient almost vanishes 
within the bubble.

(ii) Second, for  the case  $\Delta T=0.054T_{\rm c}$, Fig.3 
shows droplet migration   toward the bottom. 
Fig.4 displays the velocity field in the steady state 
taken at $t=8\times 10^4$. The velocity component through the 
interface is more increased than in Fig.1, 
again resulting in 
a flat temperature  inside the droplet as in Fig.5. 
As can be seen in Figs.4 and 5,  
a thin liquid layer is sandwiched 
between  the bottom boundary and the droplet. 
The  thickness of this layer  
is  so thin that we can define 
an apparent  contact angle 
$\theta_{\rm eff}$, 
 which is a decreasing function of 
$\Delta T$.   
In accord with these results,   
Garrabos {\it et al.} 
observed  that gas spreads on a heated wall 
initially wetted by liquid and exhibits 
  an apparent contact angle \cite{Beysens}.

(iii) Third, we increased $\Delta T$ to $0.10125T_{\rm c}$, 
where the bottom temperature is $0.97 T_{\rm c}$. 
 As shown in  Fig.6,    
phase separation occurred  to produce 
a gas domain  at the bottom, into which 
the suspended droplet was finally  
absorbed.  In the steady state  in Fig.7,  
the bottom is  covered by 
a gas layer or is completely wetted by gas \cite{wet}, 
 but the interface  with 
the bulk liquid  is still curved and 
a small velocity field is induced. 
As can be seen in Fig.8  taken at $t=3\times 10^4$, 
the temperature 
gradient within the gas layer 
becomes  steeper than 
that in the liquid. 
It is well-known that a gas film 
 on a heated wall  suppresses 
 heat transfer  to the bulk liquid and, as it thickens, 
 film boiling is induced in gravity 
 \cite{Nikolayev}.

In  the steady states the velocity  is 
of order $v_{\rm O} \sim D_T \Delta T/TL$ 
in terms of  $D_T=\lambda/\rho C_p$ from $C_p \sim \Delta s$.  
As a result, the Reynolds number($\sim Pr^{-1} R\Delta T/LT$)
 is very small, 
where   the Prandtl number $Pr= v_0\eta/mD_T $  is here 
taken to be of order 1.   
In fact,   in Figs.1, 4, and 7, 
the variance $\sqrt{\av{{\bi v}^2}}$ is 
$0.12$,  $1.9$, 
and  $0.63$,   respectively, 
in units of $10^{-3}\ell/t_0$, 
where $\ell/t_0 = D_T/Pr \ell$.
In  our simulation 
 the viscous dissipation 
is very strong even in transient states, while 
in the theory in Ref.\cite{Beysens1} 
the inertia effect (vapor recoil) 
is important in bubble motion on a heated wall. 
This difference stems from the fact 
that our droplet size is very small 
if   $\ell$ is  microscopic.

In the diffuse-interface methods 
for crystal growth \cite{Cagi} 
and  thermo-capillary flow 
 \cite{Jasnow,phase1},   
the interface width  $\ell$ may be conveniently treated   as  
a free parameter.  This is because 
the phase field 
equations  yield the  moving-interface 
problems with appropriate interface boundary conditions 
as long as $\ell \ll R(=$ domain size). 
Also in fluids  we may  take 
$\ell$ as a free parameter in the range $\ell \ll R$ 
as far as we treat the steady state  
profiles (where the Marangoni effect is suppressed 
and the Stokes approximation is valid).  However, 
 to describe  rapid fluid motions, 
the normalized viscosity 
$\eta^* (\propto \eta/\ell)$ 
should  be made very small  if  $\ell$ is enlarged.

In summary, we have developed a 
 scheme to study   
phase transition dynamics with inhomogeneous temperature 
under mass and heat fluxes. 
It can serve as a phase field model 
in fluids, though we should examine its sharp-interface 
limit \cite{phase1}. 
 For one-component fluids we mention 
wetting dynamics with evaporation and 
condensation \cite{Pomeau,deGennes},
 as  exemplified here, and  
boiling or convection in gravity 
\cite{Physica,Onukibook}. 
These effects can 
in principle  be studied  in the scheme of  
thermocapillary hydrodynamics 
supplemented with  interface conditions.  However,  
hydrodynamic calculations accounting for 
latent heat transport  
have rarely been performed.

This work is supported by 
Grant for the 21st Century COE project 
(Center for Diversity and Universality in Physics)
 from the Ministry of Education, 
Culture, Sports, Science and Technology of Japan.

\end{multicols}

\newpage 
%%%%%  Fig.1 %%%%%%%%%%%%%%%%%%
\begin{figure}[t]
\epsfxsize=2.8in
\centerline{\epsfbox{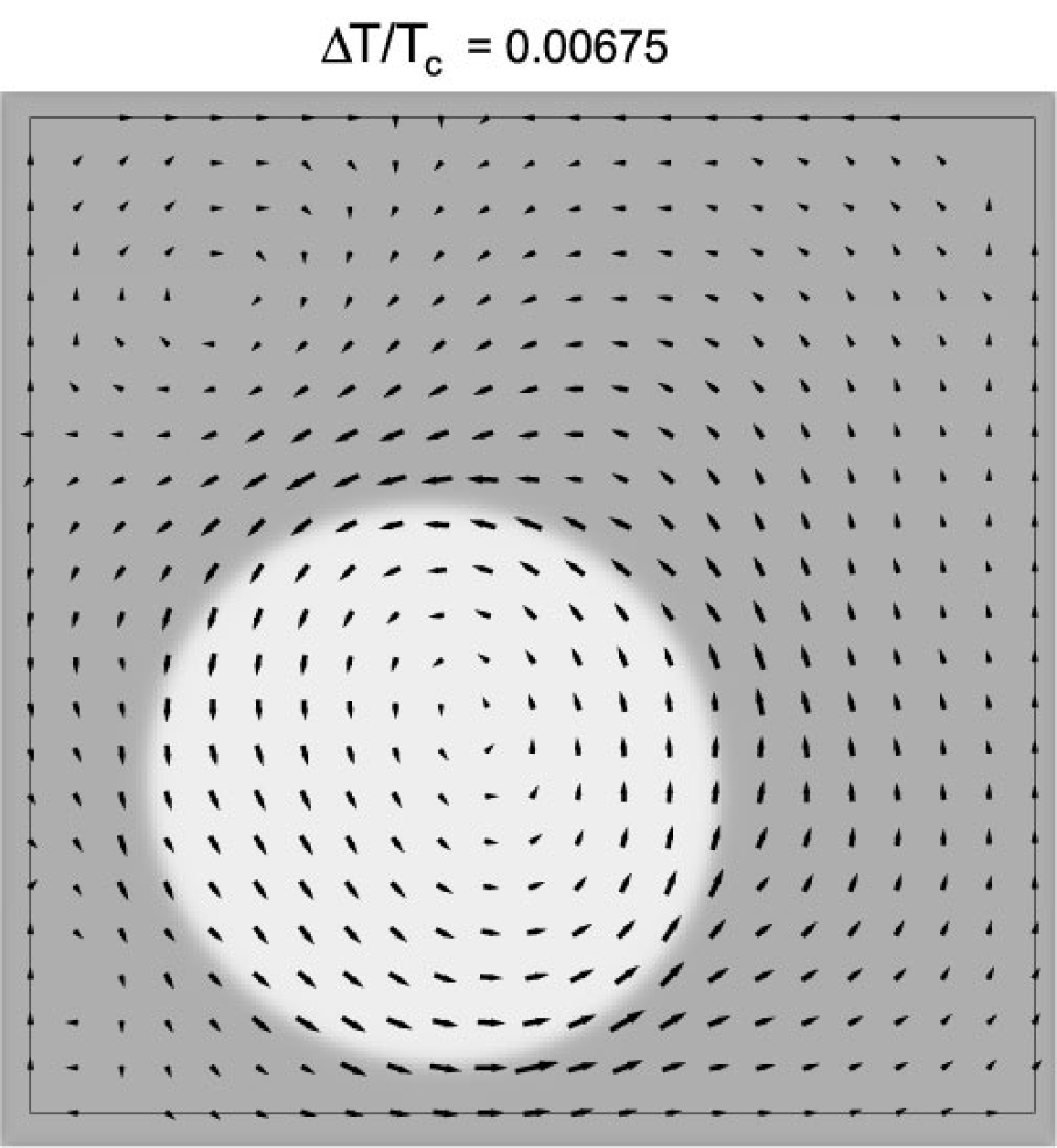}}
\caption{\protect%\narrowtext
Suspended gas droplet in a steady state in heat flow 
with $\Delta T=0.00675T_{\rm c}$. 
The arrows indicate the velocity.  
}
\label{f1}
\end{figure}
%%%%%  Fig.2 %%%%%%%%%%%%%%%%%%
\begin{figure}[t]
\epsfxsize=4.4in
\centerline{\epsfbox{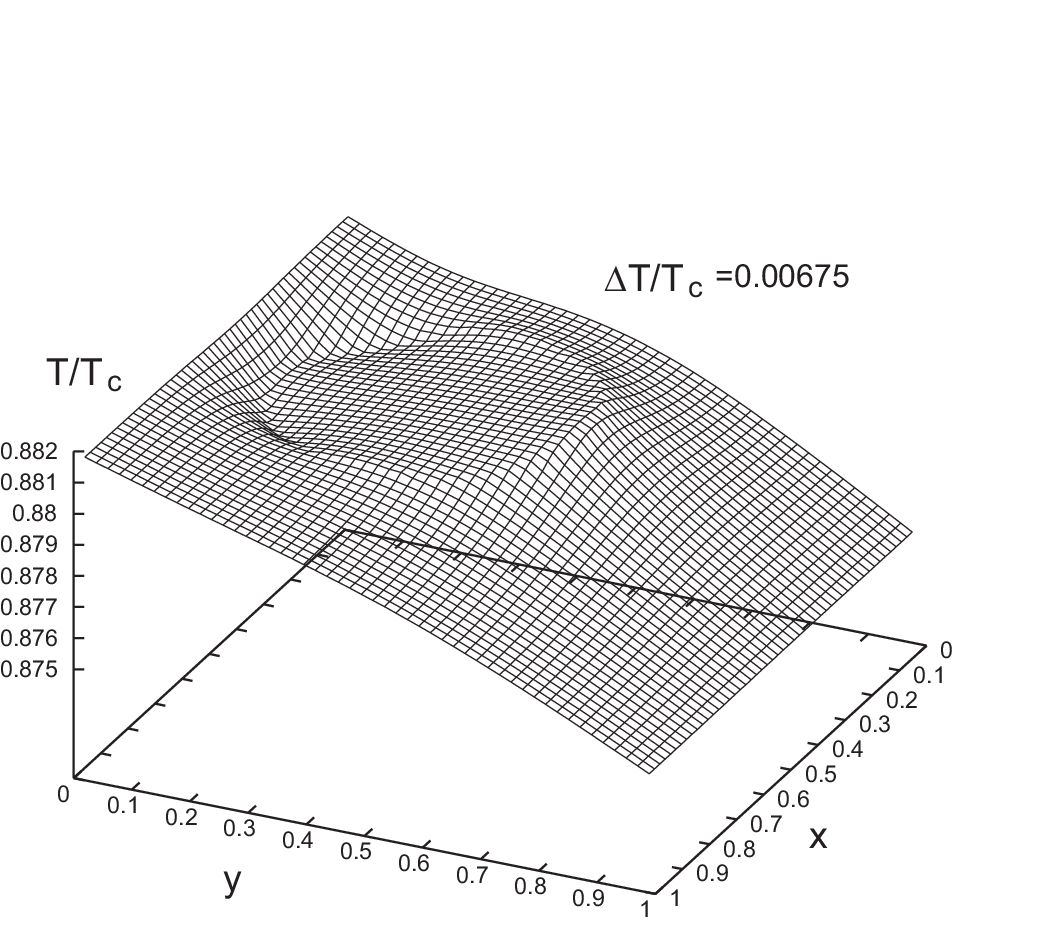}}
\caption{\protect%\narrowtext
The temperature in the steady state in Fig.1. 
It is flat within the 
gas droplet due to latent heat transport. 
}
\label{f2}
\end{figure}
%%%%% Fig.3 %%%%%%%%%%%%%%%%%%
\begin{figure}[t]
\epsfxsize=3.4in
\centerline{\epsfbox{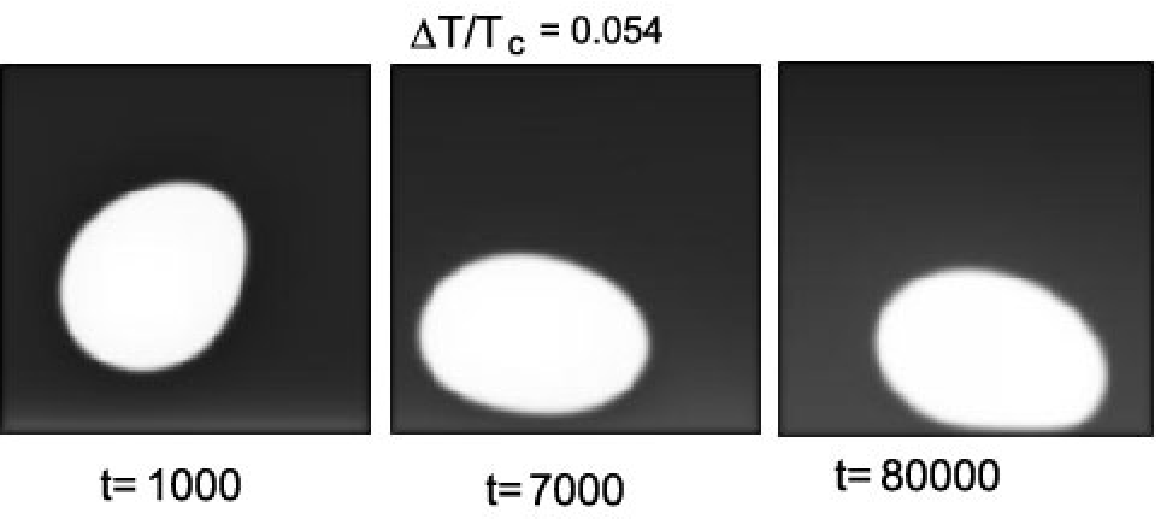}}
\caption{\protect%\narrowtext 
Migration of a gas droplet 
for $\Delta T=0.054 T_{\rm c}$. 
It   apparently 
wets the bottom  partially in the steady state 
(left), where  a thin liquid 
layer is still sandwiched. }
\label{f3}
\end{figure}

%%%%% Fig.4%%%%%%%%%%%%%%%%%%
\begin{figure}[t]
\epsfxsize=2.8in
\centerline{\epsfbox{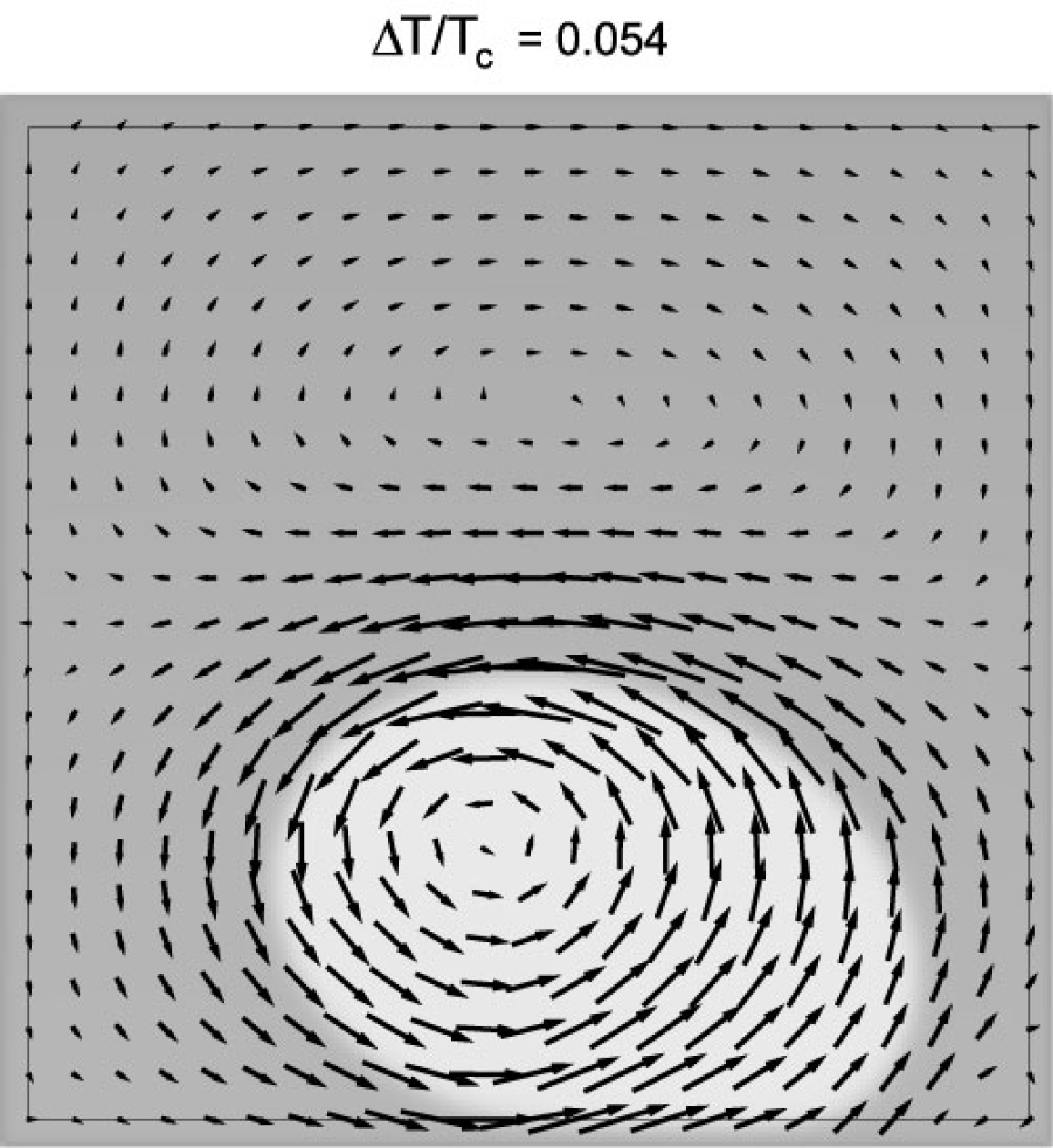}}
\caption{\protect%\narrowtext 
The velocity field around the gas droplet at 
$t=8\times 10^4$ in Fig.3.}
\label{f4}
\end{figure}

%%%%% Fig.5 %%%%%%%%%%%%%%%%%%
\begin{figure}[t]
\epsfxsize=3.4in
\centerline{\epsfbox{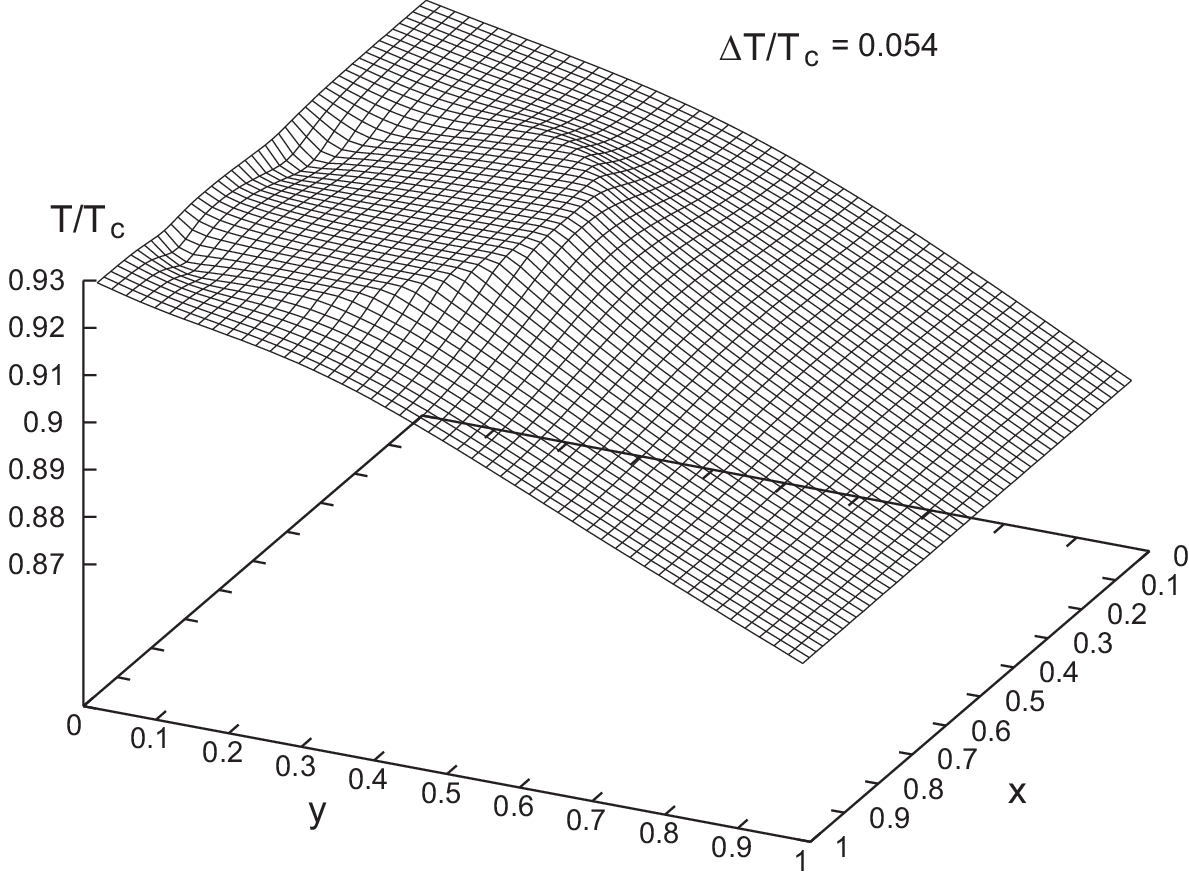}}
\caption{\protect%\narrowtext 
The temperature 
in the nearly steady state in Fig.4.}
\label{f5}
\end{figure}

%%%%%  Fig.6 %%%%%%%%%%%%%%%%%%

\begin{figure}[t]
\epsfxsize=3.4in
\centerline{\epsfbox{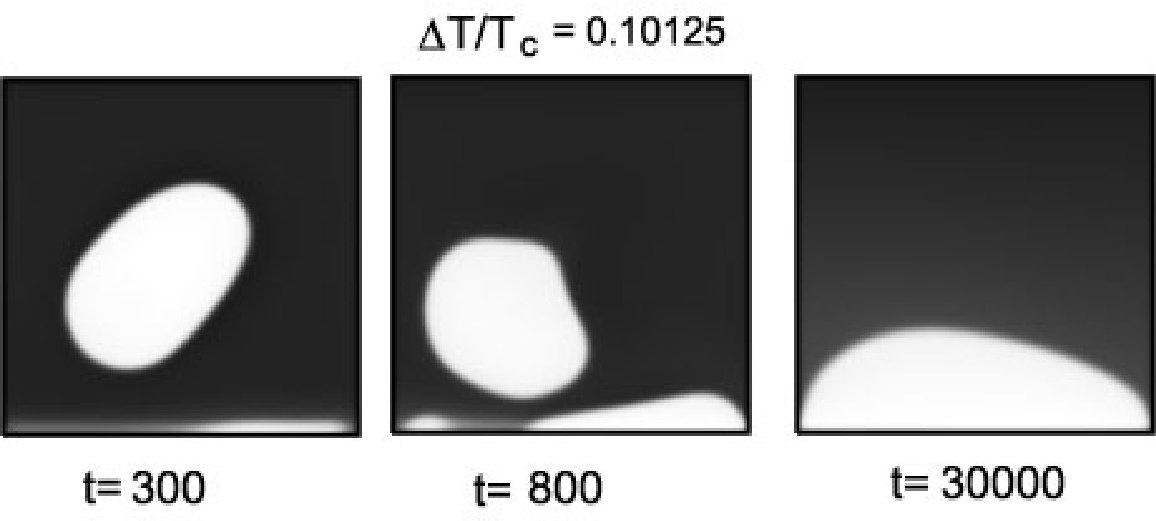}}
\caption{\protect%\narrowtext 
Time evolution for  $\Delta T=0.10125 T_{\rm c}$.
Phase separation occurs at the bottom. 
The bottom is completely wetted 
by gas in the steady state.}
\end{figure}

%%%%%  Fig.7 %%%%%%%%%%%%%%%%%%
\begin{figure}[t]
\epsfxsize=2.8in
\centerline{\epsfbox{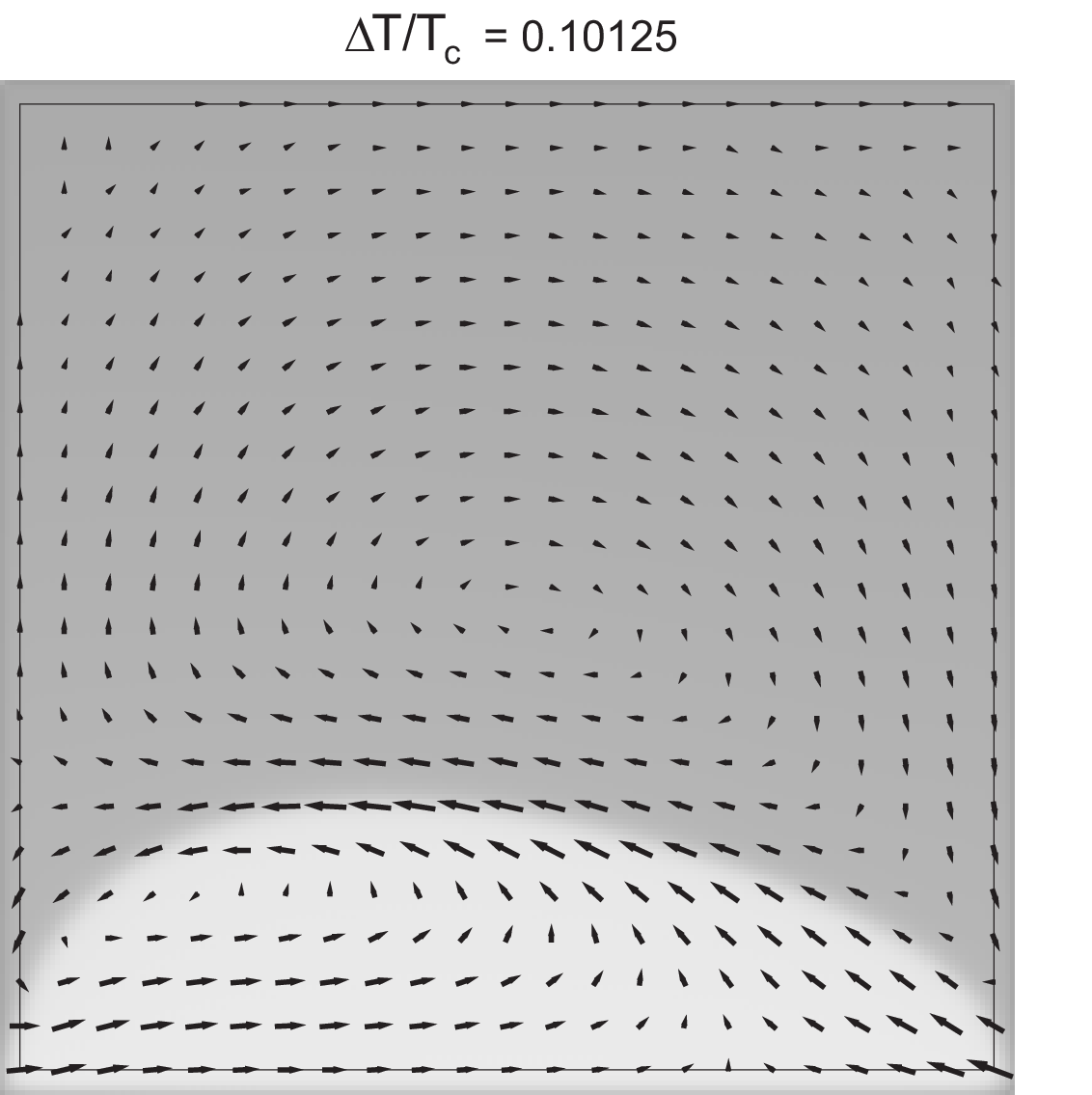}}
\caption{\protect%\narrowtext 
The velocity field around 
the gas region at $t= 3\times 10^4$ 
in Fig.6.}
\end{figure}

%%%%%  Fig.8  %%%%%%%%%%%%%%%%%%
\begin{figure}[t]
\epsfxsize=3.4in
\centerline{\epsfbox{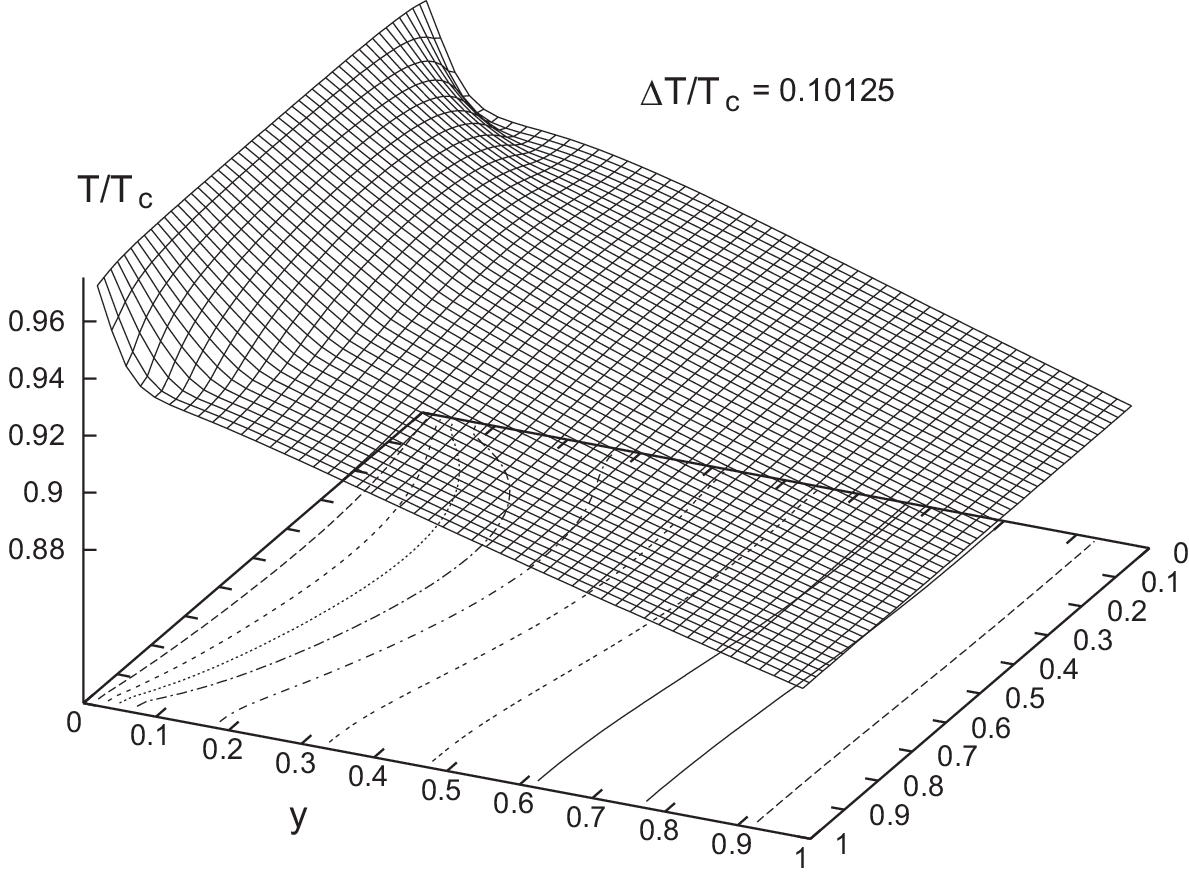}}
\caption{\protect%\narrowtext 
The temperature in the nearly 
steady state in Fig.7.}
\end{figure}


\begin{references}
\bibitem{OnukiM} A. Onuki, to be published. 
The temperature gradient inside a droplet 
is smaller that that far from it by 
$1/M_1M_2$, where 
$M_2= R\rho'\Delta s/\gamma_1$ 
is  also of order $R/a@\gg 1$.
\bibitem{wet} 
At large heat flux 
the density steeply  changes 
within a few lattice points  at 
the heated wall. Here  the boundary condition Eq.(16) is  
questionable.
\end{references}
\end{document}